\documentclass[conference]{IEEEtran}
\IEEEoverridecommandlockouts
\usepackage{cite}
\usepackage{amsmath,amssymb,amsfonts}
\usepackage{algorithmic}
\usepackage{graphicx}
\usepackage{textcomp}
\usepackage{xcolor}
\usepackage{array}
\usepackage{multirow}
\usepackage{array}
\usepackage{float}
\usepackage{tikz}
\usepackage{hyperref}

\newcommand{\specialcell}[2][c]{%
    \begin{tabular}[#1]{@{}c@{}}#2\end{tabular}}
    
\usepackage{xcolor}
\def\BibTeX{{\rm B\kern-.05em{\sc i\kern-.025em b}\kern-.08em
    T\kern-.1667em\lower.7ex\hbox{E}\kern-.125emX}}

\makeatletter



\title{Retinal Image Restoration and Vessel Segmentation using Modified Cycle-CBAM and CBAM-UNet}

\begin{document}

\author{\IEEEauthorblockN{Alnur Alimanov}
\IEEEauthorblockA{\textit{Department of Computer Engineering} \\
\textit{Bahcesehir University, Istanbul, Turkey}\\
Email: alnur.alimanov@bahcesehir.edu.tr}
\and
\IEEEauthorblockN{Md Baharul Islam}
\IEEEauthorblockA{\textit{Bahcesehir University, Istanbul, Turkey} \\
\textit{American University of Malta}\\
ORCID: 0000-0002-9928-5776}

}


\maketitle

\IEEEpubidadjcol







\begin{abstract}
Clinical screening with low-quality fundus images is challenging and significantly leads to misdiagnosis. This paper addresses the issue of improving the retinal image quality and vessel segmentation through retinal image restoration. More specifically, a cycle-consistent generative adversarial network (CycleGAN) with a convolution block attention module (CBAM) is used for retinal image restoration. A modified UNet is used for retinal vessel segmentation for the restored retinal images (CBAM-UNet). The proposed model consists of two generators and two discriminators. Generators translate images from one domain to another, i.e., from low to high quality and vice versa. Discriminators classify generated and original images. The retinal vessel segmentation model uses downsampling, bottlenecking, and upsampling layers to generate segmented images. The CBAM has been used to enhance the feature extraction of these models. The proposed method does not require paired image datasets, which are challenging to produce. Instead, it uses unpaired data that consists of low- and high-quality fundus images retrieved from publicly available datasets. The restoration performance of the proposed method was evaluated using full-reference evaluation metrics, e.g., peak signal-to-noise ratio (PSNR) and structural similarity index measure (SSIM). The retinal vessel segmentation performance was compared with the ground-truth fundus images. The proposed method can significantly reduce the degradation effects caused by out-of-focus blurring, color distortion, low, high, and uneven illumination. Experimental results show the effectiveness of the proposed method for retinal image restoration and vessel segmentation. The codes are available at \url{https://github.com/AAleka/Cycle-CBAM-and-CBAM-UNet}.
\end{abstract}


\section{Introduction}

\begin{figure}[h!]
    \centering
    \includegraphics[width=0.15 \textwidth]{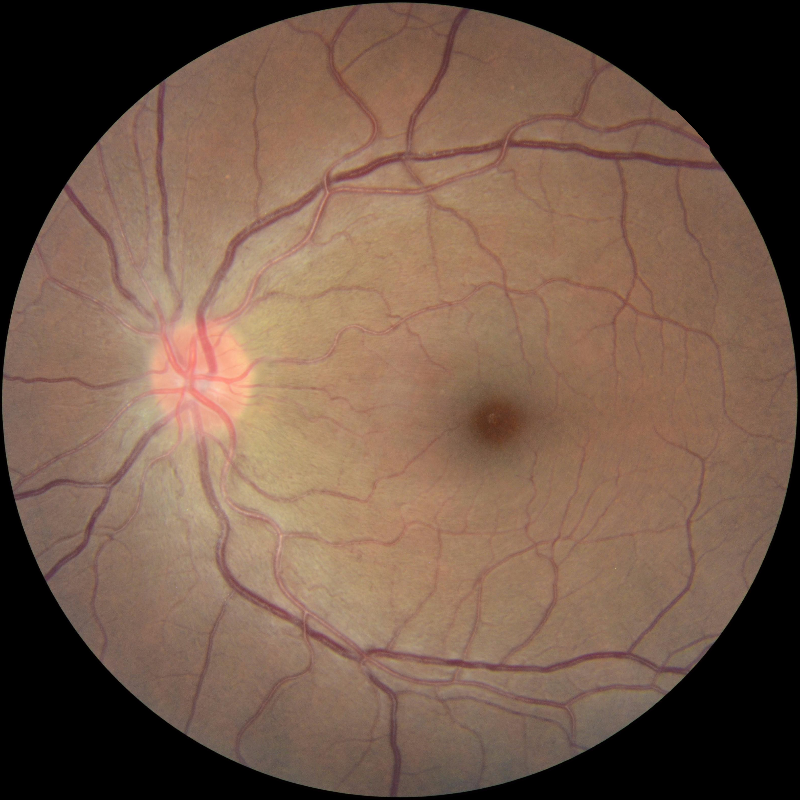}
    \includegraphics[width=0.15 \textwidth]{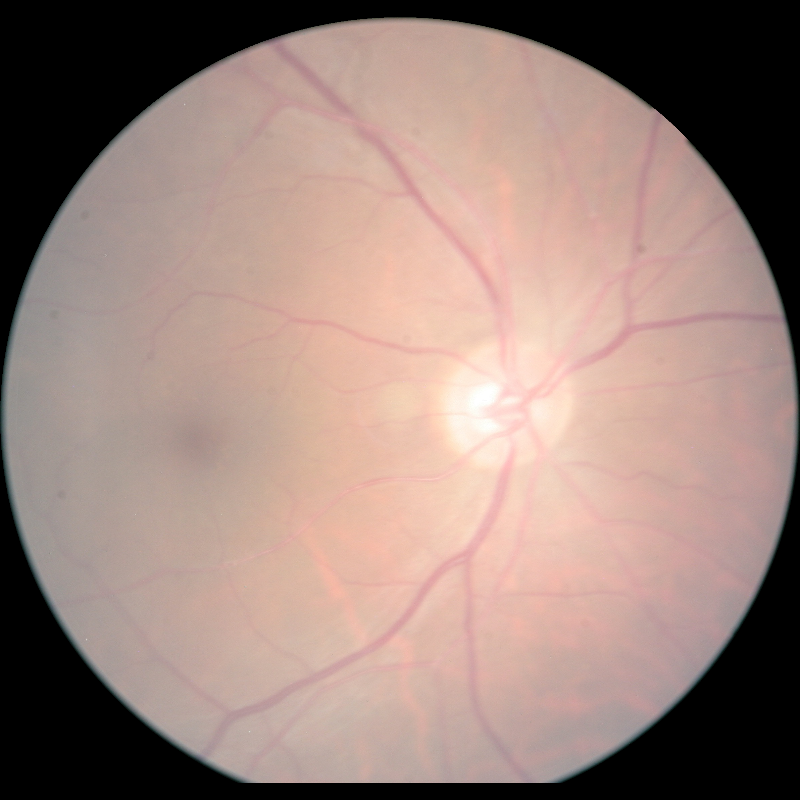}
    \includegraphics[width=0.15 \textwidth]{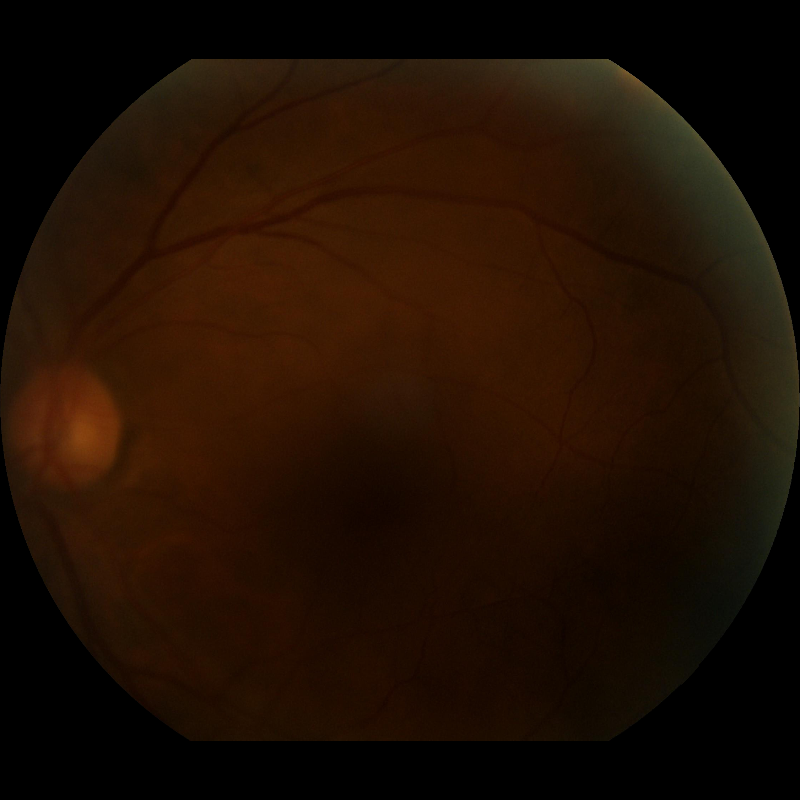}
    
    \vspace{0.1cm}
    \includegraphics[width=0.15 \textwidth]{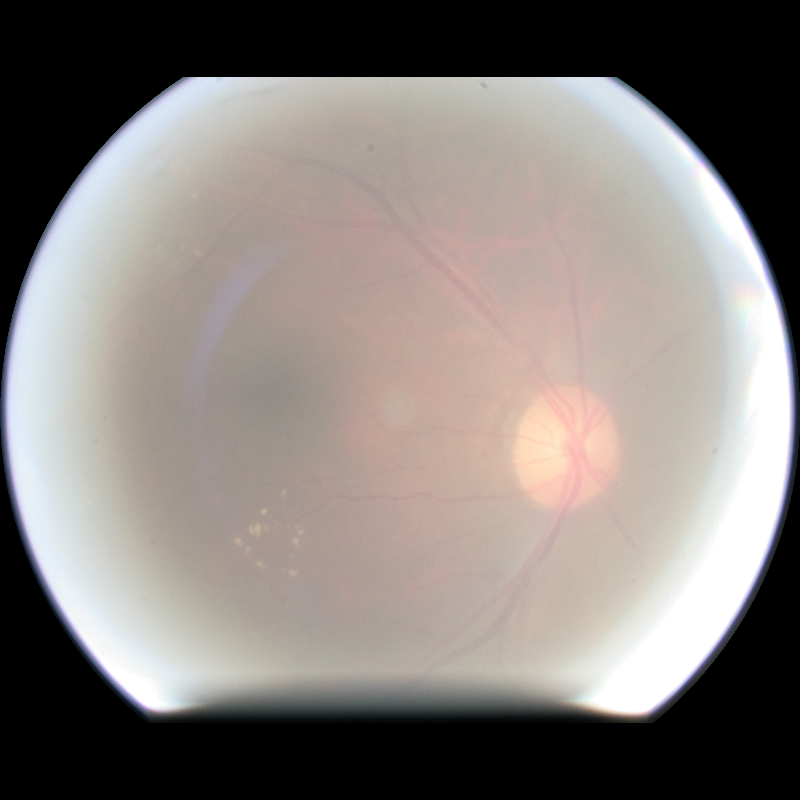}
    \includegraphics[width=0.15 \textwidth]{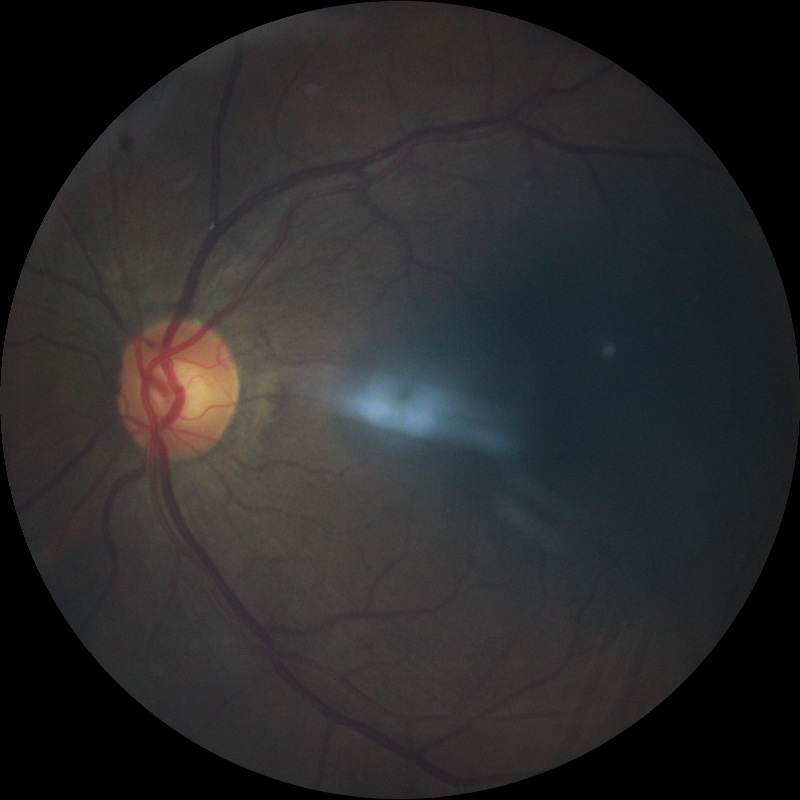}
    \includegraphics[width=0.15 \textwidth]{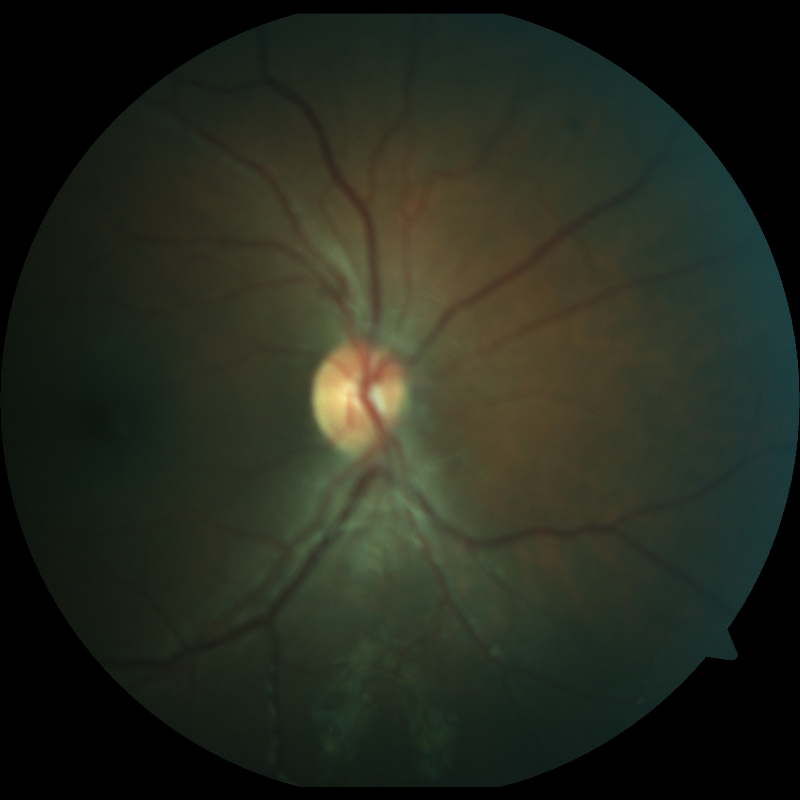}    
    \caption{Retinal image qualities. Top-bottom, left-right: original high-quality image, blurring, low, high, uneven illumination, and color distortion.}
    \label{fig:quality}
\end{figure}

To analyze human organs, body extremities, and other tissues, experts use medical imaging technologies such as magnetic resonance imaging, ultrasound, computed tomography, X-ray, and retinal imaging technology. However, images acquired from these imaging modalities may suffer from poor quality, usually caused by blurriness, noise, illumination problems, and image artifacts. Overcoming these obstacles would help with better visual perception, understanding, and analysis. Therefore, various image reconstruction techniques have been proposed using deep learning.

High-quality retinal images allow to diagnose various eye diseases, e.g., glaucoma, diabetic retinopathy, retinal tear and detachment, macular hole, and degeneration \cite{das2018survey}. However, these high-quality retinal images are challenging to acquire due to costly equipment and misoperations during medical imaging. According to a screening study \cite{philip2005impact}, about 11.9\% percent of the generated fundus images are of poor quality. The main degradation types of retinal images are blurriness, low, high, and uneven illumination, and color distortion, as demonstrated in Fig. \ref{fig:quality} as an example. As a result, ophthalmologists or experts must repeat the procedure until high-quality fundus images are produced.

\begin{figure*}[tbh]
    \centering
    \includegraphics[width=0.81\textwidth]{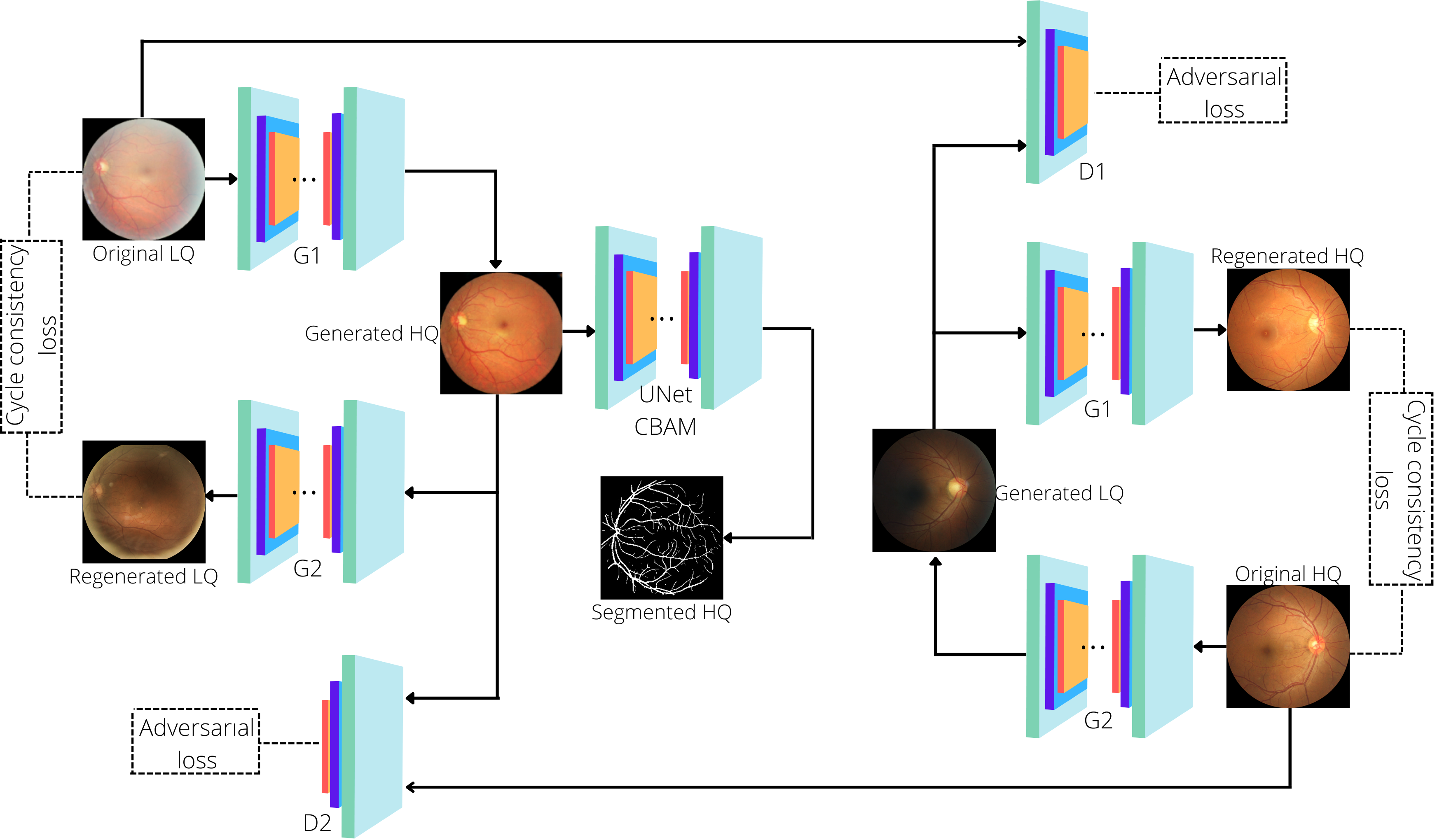}
    \caption{The workflow of the proposed method. The Original LQ image is first inputted into Generator $G_{1}$ and the output (Generated HQ) is used as input for Generator $G_{2}$ and Discriminator $D_{2}$. Next, the Original HQ image is used to make Generated LQ image using Generator $G_{2}$, then this generated image is passed to both Discriminator $D_{1}$ and Generator $G_{1}$. After these steps, cycle consistency and adversarial losses are calculated. During testing, the Generataed HQ image is then segmented using our CBAM-UNet.}
    \label{fig:dataflow}
\end{figure*}

The state-of-the-art methods can be divided into traditional hand-crafted and machine-learning-based methods. Traditional methods \cite{datta2015new, shamsudeen2016enhancement, xiong2017enhancement} are mainly based on luminosity and contrast normalization. However, these methods are challenging to generalize to all cases. For example, Shamsudeen et al. \cite{shamsudeen2016enhancement} used contrast-limited adaptive histogram equalization (CLAHE) to improve the luminosity and contrast of retinal images. However, it suffers from color distortion and additional noise. In recent years, deep learning-based methods \cite{wan2021retinal, biswas2020dvae, li2022annotation, zhao2019data, perez2020conditional, sengupta2020desupgan, you2019fundus, shen2020modeling, yoo2020cyclegan} have been introduced to perform similar tasks with high efficiency. These methods significantly outperform the traditional methods because of their generalization ability. However, these methods apply manual degradation to retrieve paired low-quality and high-quality images. The manual degradation is fundamentally different from the original. For instance, Biswas et al. \cite{biswas2020dvae} degraded the high-quality retinal images by adding noise and blurriness to retrieve image pairs. Then, they used a variational autoencoder to enhance retinal images. 

Therefore, we present a retinal image restoration method based on a cycle-consistent generative adversarial network, Cycle-CBAM, that relies on unpaired low-quality and high-quality retrieved images before proposing the retinal vessel segmentation method CBAM-UNet. The Cycle-CBAM does not rely on manually degraded images, making it superior to other methods. It also trains the model to preserve the structural details of fundus images. For the vessel segmentation task, we propose the CBAM-UNet that was trained and tested with three publicly available datasets, such as DRIVE \cite{staal2004ridge}, CHASE DB1 \cite{fraz2012ensemble}, and STARE \cite{hoover2000locating}. The main contributions of this paper are summarised below:

\begin{itemize}
    \item Due to the lack of paired low- and high-quality retinal images, the existing works focused on manually degrading high-quality images and used a small number of images. Therefore, we present a model for retinal image restoration and vessel segmentation without requiring the degraded pair of images.
    \item The proposed method eliminates the need for paired low- and high-quality images to train the network, considerably reducing the difficulty of image collection.
    \item The discriminator architecture has been significantly modified to improve the performance compared to the state-of-the-art methods.
    \item We modified UNet by changing its architecture and embedding CBAM, which led to better quantitative and qualitative results.
\end{itemize}

\section{Methodology}

\subsection{Cycle-Consistent GAN Architecture}

In this paper, we used CycleGAN that was firstly proposed in \cite{zhu2017unpaired}. The architecture of this model is shown in Fig. \ref{fig:dataflow}, it consists of two generators ($G_1$ and $G_2$) and two discriminators ($D_1$ and $D_2$). The goal of $G_1$ is to generate fake high-quality images from corresponding low-quality ones, whereas $G_2$ generates fake low-quality images from high-quality ones. When fake images are generated, $D_1$ tries to distinguish real low-quality image from fake one and $D_2$ chooses between real high-quality and fake high-quality. In that way, the model learns to translate low-quality images to high-quality ones and vice versa.

\subsection{Generators}

Generators $G_1$ and $G_2$ are fully convolutional networks, that consist of 3 down-sampling, 9 residual blocks with convolution block attention module (Res-CBAM), and 3 up-sampling blocks. Each down-sampling block consists of 2D convolution, instance normalization, and ReLU activation. The numbers of filters in convolution layers are 64, 128, and 256, respectively. Kernel sizes and strides are (7, 7), (3, 3), (3, 3) and (1, 1), (2, 2), (2, 2), respectively. Each Res-CBAM block contains 2D convolutions with 256 filters having a same kernel size of (3, 3), and strides of (1, 1) and CBAM.. The CBAM is capable of sequentially inferring attention maps along two channel and spatial dimensions from an intermediate feature map, and then multiplying the attention maps to the input feature map for adaptive feature refinement. Finally, there are 3 up-sampling blocks that consist of 2D transpose convolution, instance normalization, and ReLU activation for the first two blocks and tanh activation for the last block. Transpose convolution layers have 128, 64, and 3 filters and (3, 3), (3, 3), and (7, 7) kernel sizes, respectively. The padding values are (2, 2), (2, 2), and (1, 1).


\subsection{Discriminators}

Discriminators $D_1$ and $D_2$ contain 2D convolution layer with LeakyReLU activation, 4 down-sampling blocks, and an output 2D convolution layer. The filter values are 64, 128, 256, 512, 512, and 1, respectively. The kernel sizes are (4, 4) for all layers and stride values are (2, 2), (2, 2), (2, 2), (2, 2), (1, 1), (1, 1), respectively. They are based on a PatchGAN network \cite{li2016precomputed} that aims to classify 70x70 patches of images instead of the whole image. 


\subsection{Loss functions}

\subsubsection{Adversarial loss}

To train this network a combination of 2 loss functions was used. The first loss function is adversarial loss that was proposed in \cite{goodfellow2014generative}. $G_1$ aims to generate fake high-quality images \^{Y} from given low-quality ones X (i.e. \^{Y}=$G_1$(X)), while $G_2$ tries to convert high-quality images Y to fake low-quality ones \^{X} (i.e. \^{X} = $G_2$(Y)). After that, discriminator $D_1$ receives two images X and \^{X}, while discriminator $D_2$ gets \^{Y} and Y as inputs. $D_1$ and $D_2$ predict if the given patches of these image are real or fake using mean square error. The formulas for the loss function are expressed below:
\begin{equation}
    L_{ADV}(D_{1}, X, \text{\^{X}}) = E_{X}[log(D_{1}(X))] + E_{\text{\^{X}}}[log(1 - D_{1}(\text{\^{X}}))] 
\end{equation}
\begin{equation}
    L_{ADV}(D_{2}, Y, \text{\^{Y}}) = E_{Y}[log(D_{2}(Y))] + E_{\text{\^{Y}}}[log(1 - D_{2}(\text{\^{Y}}))]
\end{equation}

\noindent where generators are trying to minimize this value and discriminators trying to maximize it.

\subsubsection{Cycle consistency loss}
According to cycle consistency, image generated from one domain to another should be easily translated back to its original domain without significant changes. Translation of X to \text{\^{Y}} and then back to \text{\^{X}} is called forward cycle consistency, whereas Y $\rightarrow$ \text{\^{X}} $\rightarrow$ \text{\^{Y}} is called backward cycle consistency. Cycle consistency loss can be expressed in the following form:
\begin{equation}
    L_{cyc}(X,  \text{\^{X}}, Y, \text{\^{Y}}) = E_{X}[||X - \text{\^{X}}||_{1}] + E_{Y}[||Y - \text{\^{Y}}||_{1}]
\end{equation}

\subsubsection{Full loss}
The objective loss function is shown above:
\begin{align}
    L(G_{1}, G_{2}, D_{1}, D_{2}, X,  \text{\^{X}}, Y, \text{\^{Y}}) = L_{ADV}(D_{1}, X, \text{\^{X}}) + \nonumber \\
    + L_{ADV}(D_{2}, Y, \text{\^{Y}})+ \\ 
    + \lambda L_{cyc}(G_{1}, G_{2}) \nonumber
\end{align}

\noindent where $\lambda$ is the weight of reconstruction and discriminant losses. During training, the model's main goal is to solve the following problem:
\begin{equation}
    G_{1}^{*}, G_{2}^{*} = arg\min_{G_{1}, G_{2}}\max_{D_{1}, D_{2}}L(G_{1}, G_{2}, D_{1}, D_{2})
\end{equation}



\subsection{Vessel Segmentation Network}

In the medical analysis of fundus images, vessels play a significant role. By analyzing these vessels' length, width, and curvature, ophthalmologists can detect various eye diseases. Therefore, the vessel segmentation task is crucial in eye care and screening. Some vessels are short and thin, so it becomes very challenging to segment them, especially if the fundus image is not of high quality. We propose a modified UNet initially developed by Ronneberger et al. \cite{ronneberger2015u}. It adopts an encoder-decoder structure, where the encoder consists of 4 down-sampling blocks with two convolution layers, instance normalization, ReLU activation function, convolution block attention module, and a max-pooling layer. The decoder contains 4 upsampling blocks with one convolution, concatenation, and two more convolution layers with ReLU activation function, instance normalization, and CBAM. The model was trained and tested using three datasets with an Adam optimizer with a learning rate of 0.0001. 

\begin{figure}[htb]
    \centering
    \includegraphics[width=0.45\textwidth]{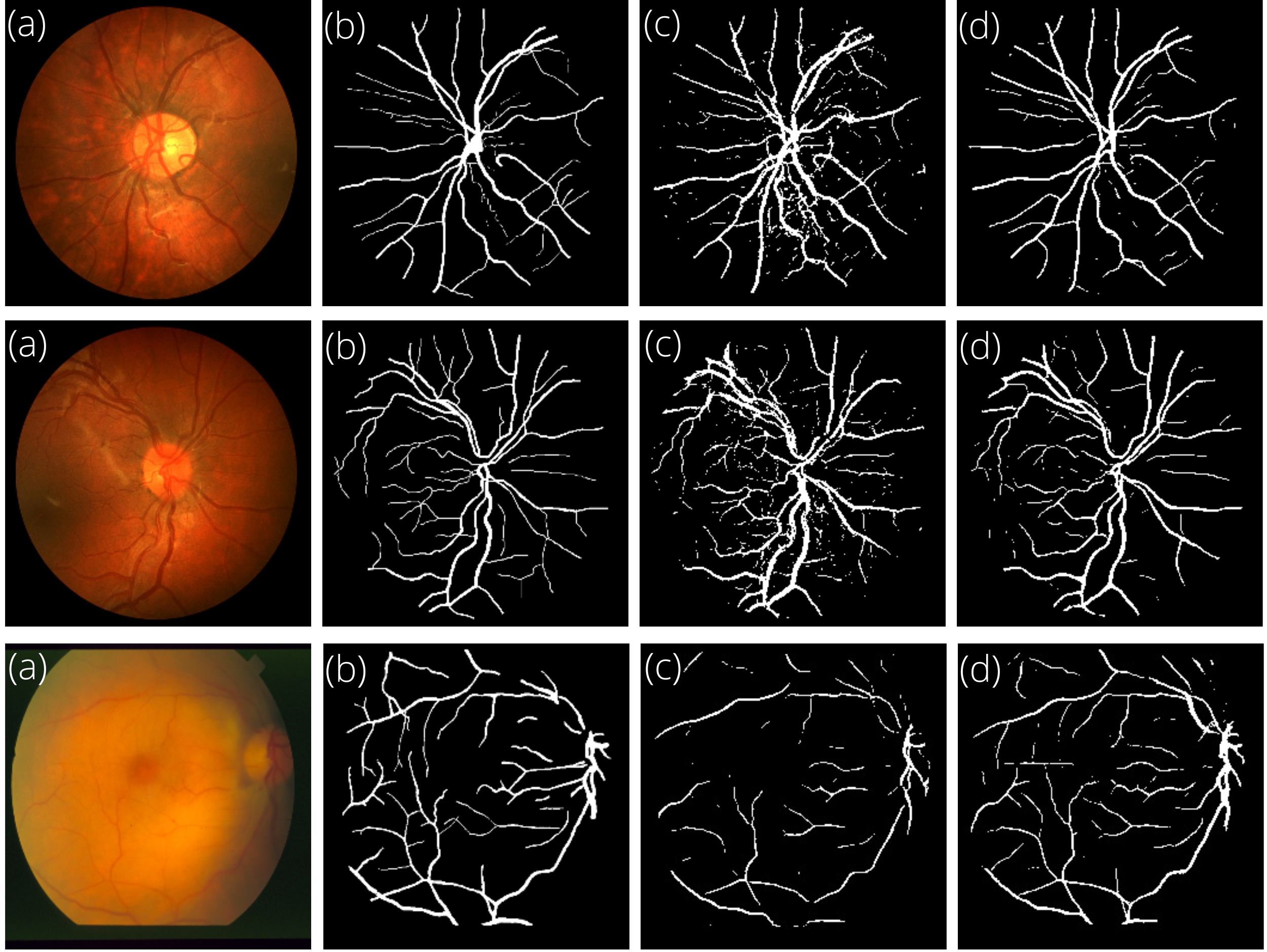}
    \caption{Qualitative analysis of proposed CBAM-UNet. Here, (a) is the original fundus image, (b) is manually segmented fundus image (ground truth), (c) is the result of UNet \cite{ronneberger2015u}, and (d) is the result of our CBAM-UNet.}
    \label{fig:cbamunet_qualitative}
\end{figure}

\begin{figure*}[htb]
\centering
    \includegraphics[width=0.9\textwidth]{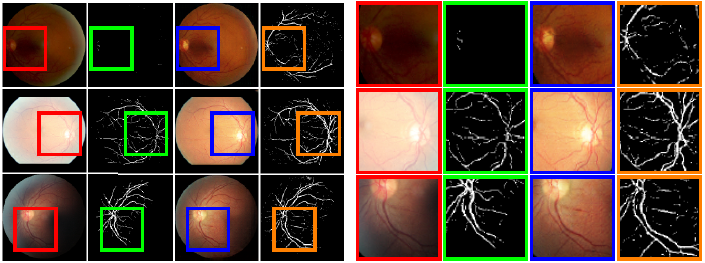} 
    \caption{Original low, high, and uneven illuminations and corresponding restored high-quality images with respective vessel segmentations.}
    \label{fig:illuminations}  
\end{figure*}

\begin{figure}[htb]
\centering
    \includegraphics[width=0.24\textwidth]{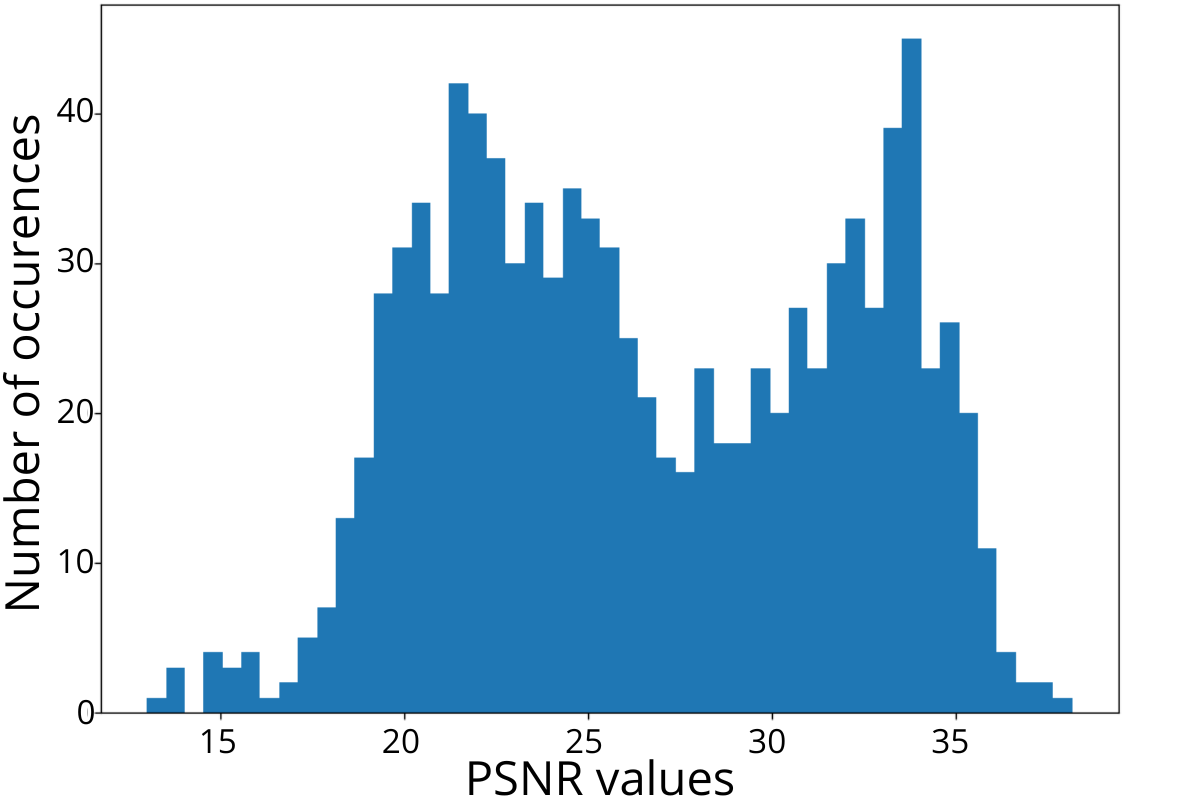}
    \includegraphics[width=0.24\textwidth]{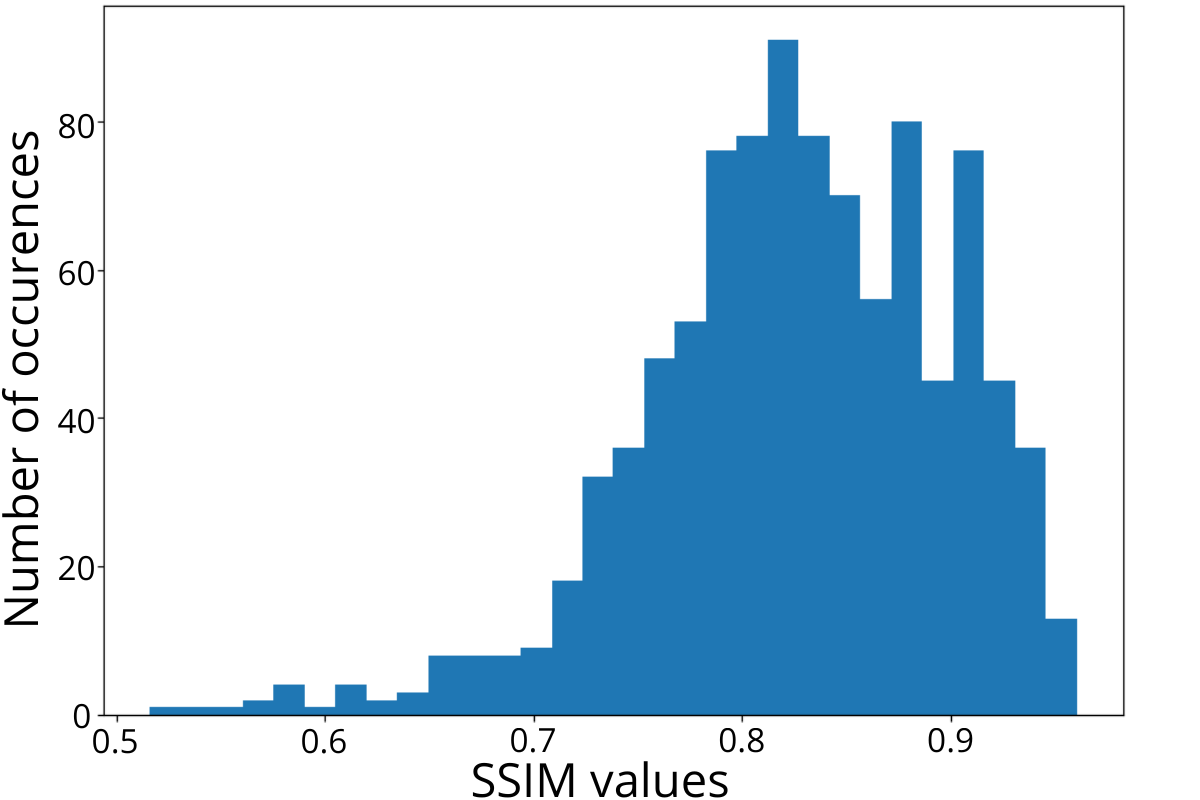}
    \caption{PSNR and SSIM values that were acquired after testing. These bar charts represent the distribution of these values.}
    \label{fig:testing}  
\end{figure}

\begin{table}[htb]
\caption{Comparison of enhancement method with other works.}
\centering
\scalebox{1}{
\begin{tabular}{|c|c|c|c|} 
 \hline
 Method & PSNR & SSIM & Time \\ [1ex]
 \hline
 CLAHE \cite{shamsudeen2016enhancement} & 25.300 dB & 0.714 & \textbf{0.01672s} \\ [1ex]
 \hline
 DPFR \cite{zhang2022double} & 12.446 dB & 0.345 & 0.44903s \\ [1ex]
 \hline
 ArcNet \cite{li2022annotation} & 20.922 dB & \textbf{0.875} & 0.18312s \\ [1ex]
 \hline
 ESRGAN \cite{wang2018esrgan} & \textbf{25.377 dB} & 0.522 & 0.55390s \\ [1ex]
 \hline
 Cycle-CBAM \cite{wan2021retinal} & 24.739 dB & 0.810 & 0.13748s \\ [1ex] 
 \hline
 Ours without CBAM & 25.212 dB & 0.793 & \textbf{0.12709s} \\ [1ex]
 \hline
 Ours with CBAM & \textbf{26.647 dB} & \textbf{0.827} & 0.13780s \\ [1ex]
 \hline
\end{tabular}}
\label{table:comparison}
\end{table} 

\setlength{\tabcolsep}{2pt}
\setlength{\extrarowheight}{1pt}
\begin{table}[htb]
\caption{Quantitative analysis of segmentation results.}
\centering
\scalebox{0.88}{
\begin{tabular}{|c|c|c|c|c|c|c|c|} 
 \hline
 Dataset & Method & Jaccard & F1 & Recall & Precision & Accuracy & FPS \\ [1ex]
 \hline
 \hline
 \multirow{2}{*}{\specialcell{DRIVE \\ \cite{staal2004ridge}}} & UNet \cite{ronneberger2015u} & 0.6520 & 0.7892 & 0.7786 & 0.8054 & 0.9639 & \textbf{23.279} \\ [1ex]
 \cline{2-8}
  & Ours & \textbf{0.6659} & \textbf{0.7992} & \textbf{0.7829} & \textbf{0.8213} & \textbf{0.9659} & 19.376 \\ [1ex]
 \hline
 \hline
 \multirow{2}{*}{\specialcell{CHASE DB1 \\ \cite{fraz2012ensemble}}} & UNet \cite{ronneberger2015u} & 0.5860 & 0.7388 & 0.7763 & 0.7078 & 0.9647 & \textbf{17.211} \\ [1ex]
 \cline{2-8}
  & Ours & \textbf{0.6077} & \textbf{0.7558} & \textbf{0.8088} & \textbf{0.7116} & \textbf{0.9665} & 14.693 \\ [1ex]
 \hline
 \hline
 \multirow{2}{*}{\specialcell{STARE \\ \cite{hoover2000locating}}} & UNet \cite{ronneberger2015u} & 0.6370 & 0.7769 & 0.8005 & 0.7667 & 0.9657 & \textbf{12.533} \\ [1ex]
 \cline{2-8}
  & Ours & \textbf{0.6656} & \textbf{0.7979} & \textbf{0.8010} & \textbf{0.7997} & \textbf{0.9697} & 11.417 \\ [1ex]
 \hline
\end{tabular}}
\label{table:SegmentationComparison}
\end{table} 

\begin{figure*}[htb]
\centering

    \includegraphics[width=0.885\textwidth]{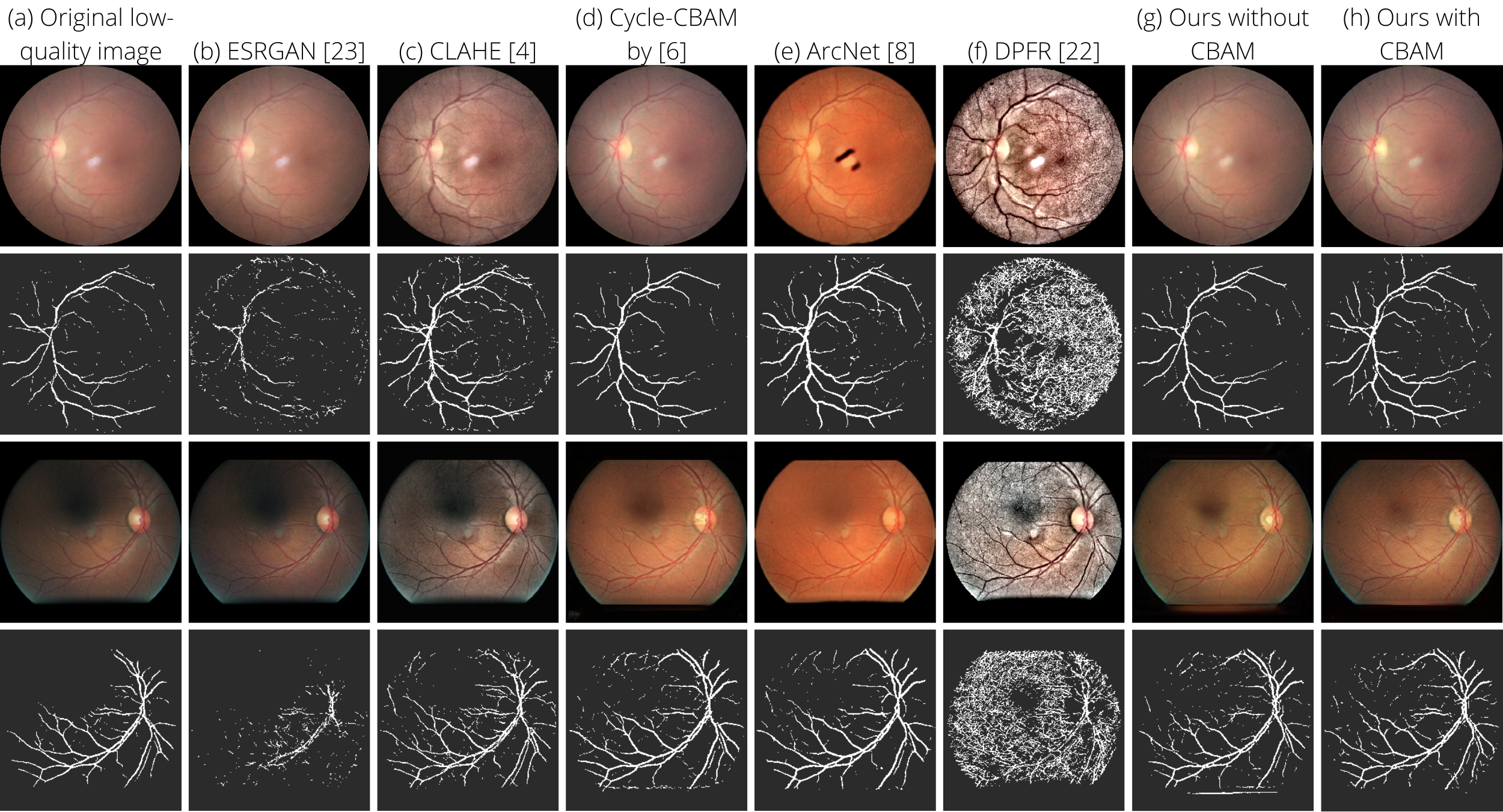} 
    
    \caption{Qualitative analysis of various methods. (a) Original low-quality image, (b) ESRGAN \cite{wang2018esrgan}, (c) CLAHE \cite{shamsudeen2016enhancement}, (d) Cycle-CBAM \cite{wan2021retinal}, (e) ArcNet \cite{li2022annotation}, (f) DPFR \cite{zhang2022double}, (g) ours without CBAM, (h) ours with CBAM.}
    \label{fig:qualitative}  
\end{figure*}

\section{Experiments}
\subsection{Restoration and Vessel Segmentation Datasets}

We used two datasets for the retinal image restoration task, such as EyeQ \cite{fu2019evaluation} and a dataset provided by Mendeley Data repositories \cite{yoo2020cyclegan}. The EyeQ dataset contains three classes of retinal images, divided according to their quality: good (16,818 images), usable (6,434 images), and bad (5,538 images). All images are fixed resolution at $800\times800$. The Mendeley dataset contains two classes: artifacts (1,146 images) and non-artifacts (1,060 images) with a resolution of $350\times346$. For training, we used "Good," "Non-artifacts," "Usable," and "Artifacts" classes. The total numbers of training high-quality and low-quality images is 17,878 and 6,441, respectively. The numbers of validation and testing images that were randomly taken from both datasets are 151 and 987, respectively. 

In addition to full-reference evaluation metrics such as PSNR and SSIM, we performed vessel segmentation for original low-quality and generated high-quality images to test the model's performance. For the vessel segmentation task, we developed a modified UNet and trained it on DRIVE \cite{staal2004ridge}, CHASE DB1 \cite{fraz2012ensemble}, STARE \cite{hoover2000locating}. The DRIVE dataset contains 40 fundus images with a resolution of $565\times584$, the CHASE DB1 dataset consists of 28 retinal images with a resolution of $999\times960$, and STARE includes a total of 20 images with a resolution of $700\times605$  with corresponding manual segmentation made by ophthalmologists. The total number of images is 88, which is insufficient for training. Therefore, we performed data augmentation using horizontal and vertical flipping and rotations. The resultant training and testing sets have 176 and 44 images, respectively.
 
\subsection{Training details}

The retinal image restoration model was implemented using the Keras framework. The hardware configuration includes an Intel Core i7-10700f CPU, 32 GB of RAM, and an NVIDIA GeForce RTX 2080 SUPER 8 GB. The model was trained for 30 epochs. Two subsets of 2,000 low-quality and 2,000 high-quality images were randomly taken from the training set at each training epoch. To evaluate the performance of the first model, PSNR and SSIM evaluation metrics have been adopted. During retinal image restoration model training, the evaluation was performed using the validation set to see the progress in every 500 training steps. This helped to know the model's performance on images not present in the training set. The validation PSNR and SSIM values had been increasing gradually at the same rate as the training PSNR and SSIM. Total training time, including validation, took 20 hours. 

The retinal vessel segmentation model was developed using the PyTorch framework. We used metrics such as Jaccard similarity coefficient, F1-score, recall, precision, and accuracy in evaluating the vessel segmentation network. The model has been set to train for 100 epochs. If the validation loss does not improve for a consecutive 5 epochs the training stops, and only the best weights are saved. As a result, the model was trained for 58 epochs.

\section{Results and Analysis}

\subsection{Qualitative analysis}
In retinal images, vessels play a significant role. Therefore, it is essential to enhance the quality and preserve the structural details of these images. Vessel segmentation has been performed using our modified CBAM-UNet. Fig. \ref{fig:cbamunet_qualitative} demonstrates the results of the original UNet \cite{ronneberger2015u} and our model along with ground truth. As we can see, our method performs better in vessel segmentation due to the ability to segment the micro parts of the retinal vessels. In addition, our method has fewer false-positive vessels, making it look clearer with much less noise. Fig. \ref{fig:illuminations} presents retinal images with various degradation types, their restored images, and the results of vessel segmentation for low-quality and restored high-quality images. The most significant difference can be noticed in low-illumination images, where blood vessels are not visible.

\subsection{Quantitative analysis}
Fig. \ref{fig:testing} displays the distributions of PSNR and SSIM values for testing. The y-axis indicates the number of occurrences, and the x-axis shows each metric's values. As we can see, most PSNR values range from 24 to 34 dB, and SSIM values are mostly above 0.8. The average value of PSNR and SSIM are 26.647 dB and 0.827, respectively. Table \ref{table:comparison} shows the PSNR and SSIM values along with time spent on a single image enhancement of our method with the state-of-the-art methods, such as Cycle-CBAM \cite{wan2021retinal}, ArcNet \cite{li2022annotation}, ESRGAN \cite{wang2018esrgan}, CLAHE \cite{shamsudeen2016enhancement} and DPFR \cite{zhang2022double}. As we can see, the proposed method outperforms other methods in PSNR, and in terms of SSIM, it comes after ArcNet \cite{li2022annotation}. Regarding time spent on single image restoration, it is outperformed by CLAHE \cite{shamsudeen2016enhancement} and ResNet.

Table \ref{table:SegmentationComparison} provides the quantitative results of the original UNet \cite{ronneberger2015u} and our modified CBAM-UNet. As we can see, the proposed network outperforms UNet in terms of all evaluation metrics, except for the FPS. On average, it takes 0.132 seconds to segment a single image.

\subsection{Comparison of restoration}
The proposed method has been qualitatively evaluated and compared with state-of-the-art methods. Fig. \ref{fig:qualitative} shows the original low-quality image along with the results of the state-of-the-art restoration methods and corresponding vessel segmentation performance. Wan et al. \cite{wan2021retinal} method generally improved the images. However, some retinal vessels are missed compared to the result of our method. ArcNet \cite{li2022annotation} produces black artifacts in some of the images. On the other hand, the CLAHE \cite{shamsudeen2016enhancement} method fails to enhance low-illumination images and increases the presence of artifacts in an image. The DPFR \cite{zhang2022double} makes the retinal structure more visible but at the same time significantly increases artifacts. ESRGAN \cite{wang2018esrgan} increased the resolution but could not enhance the details of a retinal image. Hence, it failed in vessel segmentation. CycleGAN (Fig. \ref{fig:qualitative}(g)) could enhance the images, but if we compare it to Cycle-CBAM (Fig. \ref{fig:qualitative}(h)), some vessels could not be segmented.

\begin{figure}[htb]
\centering
    \includegraphics[width=0.45\textwidth]{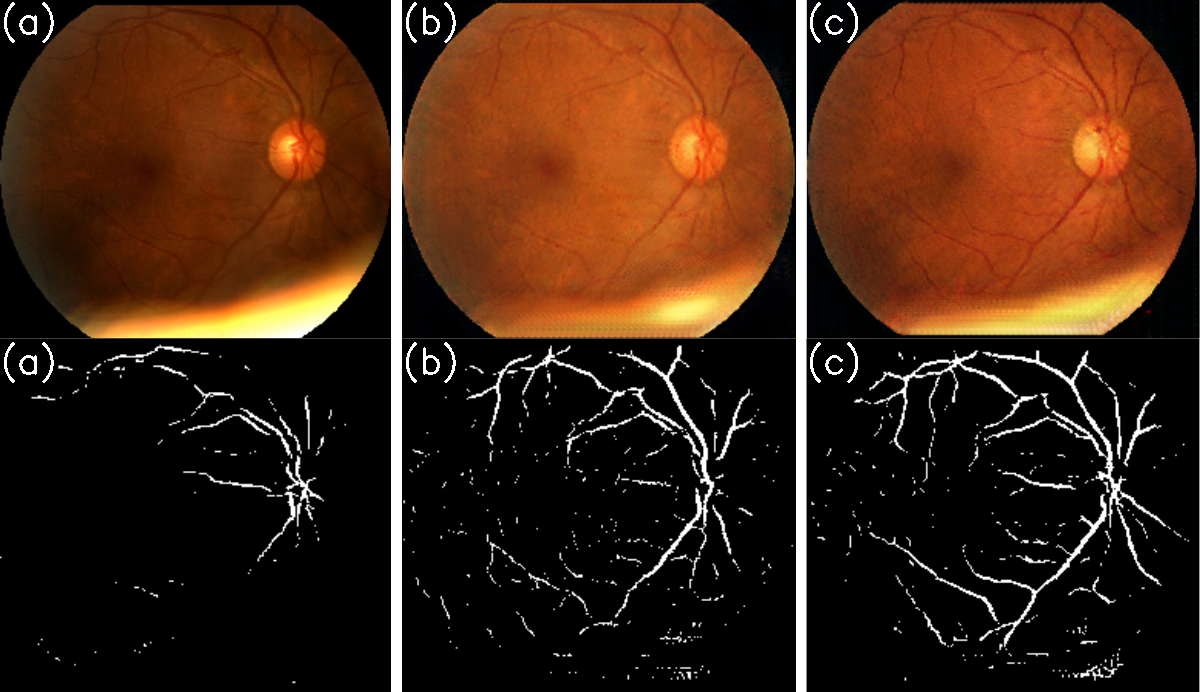} 
    \caption{Ablation study. Restoration and vessel segmentation results of the input retinal image (left) without (middle) and with (right) considering the CBAM in the model.}
    \label{fig:ablation} 
\end{figure}

\subsection{Ablation Study}
To validate the effect of the integration of CBAM into residual blocks, we tested the enhancement model with and without CBAM. The CBAM replaced ReLU activation and the convolutional layer in the residual block. As a result, the PSNR and SSIM values have been increased by 1.435 dB and 0.034, respectively. In addition, qualitative analysis has been performed to validate the efficiency of CBAM using vessel segmentation. Fig. \ref{fig:ablation} demonstrates the retinal image restoration and vessel segmentation results without (middle) and with considering the CBAM in the proposed method. The first method slightly improves the illumination problem. However, some parts of the segmented vessel are missing without considering the CBAM in the proposed method.

\section{Conclusion}
In this paper, we address the problems of low-quality retinal image restoration and vessel segmentation by proposing a cycle-consistent generative adversarial network and a retinal vessel segmentation model. The convolution block attention modules are embedded into two models to improve the feature extraction. The enhancement model does not require paired datasets because it uses unpaired low- and high-quality images. The proposed restoration and vessel segmentation methods perform better than the state-of-the-art by a large margin. Thus, it can fix the out-of-focus blurring, color distortion, and high, low, and uneven illumination problems of retinal images. However, our method may fail to restore the too low and too high luminance or uneven distribution that needs further investigation in the future.




\bibliographystyle{IEEEtran}

\bibliography{strings,refs}
\end{document}